\documentstyle[epsfig]{article}

\begin{document}

\begin{flushright}
DESY 99-172\\ 
(November 1999)
\end{flushright}

\vspace{.3in}

\begin{center}
{\Large \bf
New developments in inflationary models\footnote{To appear in the
Proceedings of the Conference "Beyond the Desert '99", Tegernsee, Germany,
6-12 June 1999.}}

\vspace{.3in}
{\large\bf  Laura Covi}

\vspace{.4 cm}
{\em DESY Theory Group,\\
Notkestrasse 85,\\
D-22603 Hamburg, Germany}

\end{center}

\begin{abstract}
We study models of inflation where the inflaton corresponds to
a flat direction in field space and its mass term is generated 
by gravity mediated soft supersymmetry breaking at high scale. 
Assuming the inflaton to have non negligible couplings to other 
fields, its mass runs with scale and can reach the small value 
required by slow roll inflation at a lower scale, even if
its initial value is too large. Slow roll inflation can therefore 
take place in such a regime, as long as the mass remains small, 
with a spectral index that is then scale dependent. We explore 
the parameter space of this kind of models to find the region 
compatible with the present observations. 
\end{abstract}

\section{Introduction}

Slow roll inflation\footnote{For a general discussion
and references on inflation see \cite{lcovi:ly98}} requires 
the flatness conditions $\epsilon\ll 1$ and $|\eta|\ll 1$ 
on the potential, where
\begin{equation}
\epsilon \equiv \frac{1}{2} M_{Pl}^2 
\left( \frac{V'}{V} \right)^2 ; \,\,\,
\eta \equiv M_{Pl}^2{V^{\prime\prime}\over V} \,,
\end{equation}
and $M_{Pl} \equiv (8\pi G)^{-1/2} = 2.4\times 10^{18}
\mbox{GeV}$.
The first condition is easily satisfied in most inflationary
models where the inflaton lies near to an extremum of the 
potential, but the second is problematic. In fact, during
inflation in a generic supergravity potential all scalar fields
\cite{lcovi:dine,lcovi:coughlan}, and in particular the inflaton
\cite{lcovi:cllsw} acquire a contribution to the mass-squared 
of magnitude $V/M_{Pl}^2$, which spoils this condition.

Very few proposals have been put forward to solve this problem 
which would not be relying on some sort of fine 
tuning~\cite{lcovi:ly98}; specific types of K\"ahler potential
and superpotential can succeed in canceling the dangerous
contribution, as it happens in no--scale supergravity.
The aim of this paper is to investigate instead the proposal of 
Stewart~\cite{lcovi:st97,lcovi:st97bis}: in this scenario 
the contribution to the inflaton mass is unsuppressed at 
high scale, but loop corrections can flatten the inflaton
potential to realize sufficient inflation without any 
significant fine-tuning.

We will explore the potential in a general model of this kind
and find the region of parameter space allowed by the observed 
magnitude and spectral index of the curvature perturbation.
We will finally discuss the naturalness of such picture
and the consequence of future observations.

\section{The running mass models}

In the model proposed by Stewart \cite{lcovi:st97}, slow-roll 
inflation occurs, with the following Renormalization Group (RG)
improved potential
for the canonically normalized inflaton field $\phi$;
\begin{equation}
V =V_0 + \frac{1}{2} m_\phi^2 (\phi) \phi^2 + 
 \frac{1}{2} m_\psi^2 (\phi) \psi^2   
 + \frac{1}{4} \lambda (\phi) \phi^2 \psi^2 
\cdots  \,.
\label{vinf}
\end{equation}
The constant term $V_0$ is supposed to come from the supersymmetry
breaking and to dominate at all relevant field values. 
Non--re\-nor\-ma\-li\-zable terms, represented by the dots, 
give the potential a minimum at large $\phi$,
but they are supposed to be negligible during inflation. 
The last two terms also vanish during inflation, since $\psi =0$, 
but are responsible for the hybrid exit from the inflationary period.

The inflaton mass-squared and all the other parameters
depend on the renormalization scale $Q$, and following 
\cite{lcovi:st97,lcovi:st97bis} we have taken $ Q=\phi $,
where now $\phi$ denotes the classical v.e.v of the inflaton field
during inflation.
Such choice for the renormalization scale minimizes the one loop
correction to the potential, since the main contribution 
goes like $\ln (\phi/Q)$ for $\phi$ larger than any other scale,
and therefore the potential in eq. (\ref{vinf}) is effectively
equivalent to the full one loop potential. 
If the inflaton v.e.v. is not the dominant scale, then some other 
choice of $Q$ will be appropriate and the simplification we have
made is no more viable. We will assume that the
inflaton v.e.v. is the dominant scale up to the end of inflation.

At the Planck scale, $m_\phi^2 (M_{Pl})$ is supposed to have
the generic magnitude
\begin{equation}
|m^2_0 |=|m^2_\phi (M_{Pl}) | \sim V_0
\label{mexpect1}
\end{equation}
coming from supergravity corrections \cite{lcovi:cllsw,lcovi:clr98}.

Without running, this would give $|\eta|\sim 1$, preventing 
slow--roll inflation. But at field values below the
Planck scale, the RG drives $m^2_\phi (\phi)$ 
to small values, corresponding to $|\eta(\phi)|\ll 1$, and slow--roll
inflation can take place. We have in fact that the slow-roll parameters
are given in our case by
\begin{eqnarray}
\epsilon &=& {M^2_{Pl}\phi^2\over 2 V_0^2}\left[ m^2_\phi (\phi) + {1\over 2}
{d m^2_\phi \over d\ln (\phi)} \right]^2\\
\eta &=& {M^2_{Pl}\over V_0}  \left[ m^2_\phi (\phi) + {3\over 2}
{d m^2_\phi \over d\ln (\phi)} + {1\over 2} {d^2 m^2_\phi \over d\ln^2(\phi)} 
\right];
\end{eqnarray}
since the derivatives of $m^2_\phi$ are suppressed by the coupling constant,
both $\epsilon$ and $\eta$ are small around the value of $\phi$ where 
the inflaton mass vanishes and inflation can successfully happen in such
conditions.

Since in this model the $\eta$ parameter changes considerably
as $\phi$ decreases, slow-roll inflation will continue until some 
epoch $\phi_{end}$, when either the critical value 
$\phi_c = - 2 m_\psi/\lambda $ is reached or $\eta(\phi)$ becomes of 
order $1$. 
To reduce the number of parameters involved in our analysis, we will 
assume the latter to be the case, so that both the flattening of the
potential and the end of slow roll inflation are due to the
mass running; the critical value $\phi_c$ will be reached after 
a brief phase of fast--roll, that should not change considerably the
e-foldings number.
We have then that the number of e-folds generated while the inflaton
runs from value $\phi$ to $\phi_{end}$ is given in the slow 
roll approximation by
\begin{eqnarray}
{\cal N}(\phi) &=& \int_{\phi_{end}}^{\phi} d\phi {V \over M^2_{Pl} V'}
\nonumber\\
&=&  {V_0\over M^2_{Pl}} \int_{\phi_{end}}^{\phi} 
{d\ln(\phi) \over m^2_\phi(\phi) + 
{1 \over 2} {d m^2_\phi(\phi)\over d\ln(\phi)}}
\label{lcovi:N}
\end{eqnarray}
where $ m^2_\phi(\phi) $ is given by solving the RG equations.

\section{The scale dependence of the spectral index}

The scale dependence of the spectral index in this kind of models
is strictly related to the RG equation of the inflaton mass.
The case of a gauge coupling dominated running has been studied
in \cite{lcovi:st97bis,lcovi:clr98,lcovi:cl98} while the Yukawa 
dominated running has been considered in \cite{lcovi:co98}. 
We will review the two extreme cases in a toy model in the 
following, concentrating in the case where inflation takes place 
while the inflaton rolls from the region $m^2_\phi \simeq 0$
towards the origin. 

A useful way to understand the general behaviour, is to 
consider the linear approximation for the running inflaton 
mass
\begin{equation}
m^2_\phi (\phi) \simeq - {V_0\over M^2_{Pl}} \left[
\mu^2_\star + c \ln (\phi/\phi_\star) \right]
\end{equation}
where the $\star $ denotes values of the variables where 
$V'$ vanishes and the constant $c$ is small and proportional 
to the relevant coupling.
We have then that the observational quantities can all be written 
as function of the three adimensional parameters
\begin{eqnarray}
c &=& - {d m^2_\phi \over d\ln(\phi)}|_{\phi=\phi_\star} 
= - 2 \mu^2_\star \\
\tau &=& - |c| \ln(\phi_\star/ M_{Pl}) \\
\sigma &=& \lim_{\phi\rightarrow\phi_\star} c e^{c {\cal N}(\phi)} 
\ln (\phi_\star/\phi),
\end{eqnarray}
where ${\cal N}(\phi) $ is given by eq.~(\ref{lcovi:N}). Assuming
the linear approximation to hold up to the end of inflation, such
expression simplifies to
\begin{eqnarray}
{\cal N}(\phi) &=&  - {1\over c} \int_{\phi_{end}}^{\phi} 
{ d\ln(\phi)\over \ln (\phi/\phi_\star)}\\
&=& - {1\over c} \ln \left| {\ln \left( \phi /\phi_\star\right)
\over \ln \left(  \phi_{end} /\phi_\star\right)} \right| .
\end{eqnarray}
Note that the e-folding number is inversely proportional to the
coupling (contained in $c$) so that a small coupling gives automatically
sufficient inflation.
We see also that the parameter $\sigma $ gives directly a measure
of the departure from the linear approximation at $\phi_{end}$,
since in the case this holds all the way, $\sigma $ should be given by
$ \sigma = \pm 1+c$ for $\phi \rightarrow \phi_\star$,
i.e. would be a number of order 1. In general such approximation 
breaks down well before $\phi_{end}$ and $\sigma$ can take also very 
large values.

We obtain then for the spectral index in the linear approximation:
\begin{equation}
n({\cal N})-1= 2\sigma e^{-c{\cal N}} - 2 c;
\label{lcovi:n-1}
\end{equation}
while the COBE normalization imposes a constraint on $V_0$:
\begin{equation}
{V_0^{1/2}\over M^2_{Pl}} = 5.3\times 10^{-4} |\sigma| \exp 
\left[ -{\tau\over c} - c {\cal N}_{COBE} 
- {\sigma \over c} e^{-c {\cal N}_{COBE}} \right].
\end{equation}

\begin{figure}[t!] 
\centerline{
\epsfig{file=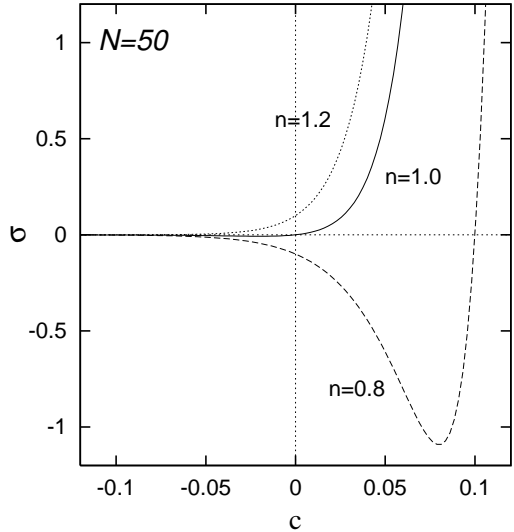,height=3in,width=3in}
}
\vspace{10pt}
\caption{
Lines of constant spectral index in the $\sigma - c$ plane
for ${\cal N} = 50$; assuming that this e-folding number
corresponds to COBE scales the allowed region is between the
long--dashed and dotted lines.
}
\label{lcovi:1}
\end{figure}

So for every particular model, the experimental constraints
limit the range  of the parameters allowed. We see in 
Fig. \ref{lcovi:1} the region in the $\sigma - c$ plane compatible
with a spectral index $|n-1| \leq 0.2 $ at ${\cal N} = 50$.

Note that every quadrant correspond to a different type of inflationary
model: 
\begin{itemize}
\item{positive $c$ implies that the inflaton mass changes sign from 
negative to positive and therefore a maximum in the RG improved potential
develops around $m^2_{\phi}\simeq 0$; in such case the sign of sigma
indicates if the inflaton is rolling towards the origin ($\sigma>0$) or 
towards large fields values ($\sigma<0$), always away from the maximum;}
\item{negative $c$ implies on the contrary that the inflaton mass changes
sign from positive to negative and a minimum in the RG improved potential
develops around $m^2_{\phi}\simeq 0$; again the sign of sigma is 
related to the direction of the inflaton's motion: $\sigma >0$ means
that the inflaton is rolling towards the minimum from the large field
values while $\sigma <0 $ from the small field values.}
\end{itemize}

The COBE normalization gives additional bounds on the value of $\tau$ 
for every choice of $c,\sigma$, since $V_0$ has to be surely larger
that the scale of nucleosynthesis (for instant reheating we have in fact
$T_{RH} = V_0^{1/4}$, but usually $T_{RH} < V_0^{1/4}$).

\section{A simple toy model}

Let us consider as an example the case of the superpotential
\begin{equation}
W = \lambda S \mbox{\rm Tr} \left(\phi_1 \phi_2\right)
\label{W}
\end{equation}
where $S$ is a singlet chiral superfield, while $\phi_i$  are chiral 
superfield
in the adjoint representation of the gauge group $SU(N)$.
We can easily compute the scalar potential given by (\ref{W})
in the limit of unbroken supersymmetry and, writing the adjoint fields in 
the fundamental basis\footnote{We define the fundamental representation
of $SU(N)$ $t_a$ such that $\mbox{\rm Tr} 
\left( t_a t_b \right) = {1\over 2} 
\delta_{ab} $ and $[t_a,t_b] = f_{abc} T_c $, while for the adjoint 
representation, e.g. $T^a_{ij} = f_{aij} $, we have 
$\mbox{\cal Tr} \left( T_a T_b \right) = N \delta_{ab} $.}
$\phi_i = \phi_i^a t_a $, it is given by:
\begin{equation}
V = {\lambda^2 \over 4} | \phi_1^a \phi_2^a|^2 +  {\lambda^2 \over 4} |S|^2 
(| \phi_1^a |^2 + |\phi_2^a |^2) + {|D_a|^2 \over 2}
\label{scalpot}
\end{equation}
where $S,\phi_i$ indicate now the scalar components of
the chiral multiplets, summation over $a$ is implicit and 
\begin{equation}
D_a = i {g\over 2} f_{abc} \left( \phi_1^{b*} \phi_1^c
+ \phi_2^{b*} \phi_2^c \right)
\label{Dterm}
\end{equation}
with $g$ denoting the $SU(N)$ gauge coupling. 

We see clearly that a flat direction exists for 
\begin{eqnarray}
S &=& 0 \\
 \phi_1^a \phi_2^a &=& 0 \\
 f_{abc} \ \phi_i^{b*} \phi_i^c &=& 0.
\end{eqnarray}
This is not the most general case and other flat directions are present,
parameterized by gauge invariant polynomials \cite{fd}. 

In the following we will consider the case when the inflaton is 
one of the components of the charged fields, i.e. we will take 
$\phi = Re \left[ \phi_1^a\right] $ to be the inflaton, while all the other 
fields are supposed to vanish during inflation. 

Then the potential for the inflaton is reduced only to the soft susy 
breaking terms \cite{ssb} and assumes the form of eq. (\ref{vinf}),
where $V_0$ is a cosmological constant that is generated by some other 
sector of the theory and is canceled in the true vacuum by the v.e.v. 
of a field in our sector, playing the role of the $\psi$. 

From supergravity, we expect all the susy breaking scalar masses,
respectively $m_S, m_{1/2}$ for the singlet and charged fields, to be of 
order of $V^{1/2}_0/M_{Pl}$ and the trilinear parameter $Y$, in 
$ {Y \over 2} \lambda S  \phi_1^a \phi_2^a + h.c.$, to be of the 
same order, as $V_0^{1/4}$ is the scale of explicit supersymmetry 
breaking during inflation. Note however that while the contribution
to the scalar masses coming from $V_0$ is always present, the trilinear
coupling $Y_0$ not always receives a contribution proportional to 
$V^{1/2}_0$.
Moreover, at the end of inflation $V_0$ vanishes and the susy breaking 
parameters will be connected instead to the gravitino mass in the
usual way, so in principle the susy breaking parameters during and
after inflation are different.

In order to write the RG improved potential, we will need to consider
the one loop renormalization group equations for all our parameters
and extract the behaviour of the inflaton mass.

Following \cite{ma93}, we write down the equations for our particle content.
The gauge field strength $\alpha = g^2/(4\pi) $ and the gaugino mass satisfy
\begin{eqnarray}
{d\alpha\over dt} &=& {\beta\over 2\pi} \alpha^2 \label{RGEalpha}\\
{d\tilde m\over dt} &=&   {\beta\over 2\pi} \alpha \tilde m
\label{RGEmtilde} 
\end{eqnarray}
where $t= \ln (Q) $ is the renormalization scale and 
$\beta = - N$ in our case of $SU(N)$ with two matter superfields in the
adjoint representation ($\beta = -3N + n_{adj} N$).

This two equations are independent from the others and their 
solution is 
\begin{eqnarray}
\alpha (t) &=& {\alpha_0 \over 1 - {\beta\over 2\pi} \alpha_0 t}
= {\alpha_0 \over 1 + \tilde \alpha_0 t}\\
\tilde m (t) &=& {\tilde m_0 \over \alpha_0} \alpha (t)
\end{eqnarray}
where $\tilde\alpha_0 = N \alpha_0/(2\pi) $ and a $0$ subscript
denotes quantities at the Planck scale.

For the Yukawa coupling, which we can always take real absorbing
its phase in the definition of the singlet field $S$, 
we have instead
\begin{equation}
{d\lambda\over dt} = - N {\alpha\over \pi} \lambda + 
{\lambda\over 16\pi^2} (N^2+1) |\lambda |^2 
\label{RGEyukawa}
\end{equation}
while for the soft susy breaking masses the equations can be
cast in a simple form using the variables
\begin{eqnarray}
m^2_{1-2} &=& m^2_1 - m^2_2, \\
m^2_{1-S} &=& m^2_1 - {1\over N^2-1} m^2_S
\end{eqnarray}
and $m^2_S$, where $m_S, m_i$ are respectively the susy breaking
masses of $S, \phi_i$ and $Y$ is the susy breaking trilinear 
coupling.
In fact we have:
\begin{eqnarray}
{dm^2_{1-2} \over dt} &=& 0 \\
{dm^2_{1-S} \over dt} &=& - {2 N \alpha\over \pi} \tilde m^2 
\label{RGE1Smass}\\
{dm^2_S\over dt} &=& {N^2 +1 \over 8\pi^2} |\lambda|^2   m^2_S + 
 {N^2 -1 \over 8\pi^2} |\lambda|^2 \left[ 2 m^2_{1-S} - m^2_{1-2} + 
{|Y|^2\over 2} \right].
\label{RGESmass} 
\end{eqnarray}

The trilinear term will have instead the equation
\begin{equation}
{dY\over dt} = {1\over 32\pi^2} (N^2+1) Y |\lambda|^2 + 
{2\over\pi} N\alpha\tilde m .
\label{RGEtrilinear} 
\end{equation}

These are a system of coupled differential equations.
We will in the next sections consider approximate solutions 
in  different cases and obtain the running inflaton mass.

\section{Dominant gauge coupling}

For $\alpha \gg \lambda^2$ a model independent analysis has been
made in \cite{lcovi:clr98}. In this case we can neglect the $\lambda^2$ 
terms and the inflaton mass running does not depend on the Yukawa 
coupling.

We have then for both charged fields
\begin{equation}
m^2_i (t)  = m^2_{i,0} - 2 \tilde m^2_0 \left[ 1- {1\over 
( 1 + \tilde \alpha_0 t)^2}\right]. 
\label{mrun1}
\end{equation}
Notice that eq.(\ref{mrun1}) gives in general a solution of 
eq.(\ref{RGE1Smass}), in the case of non negligible Yukawa.

In this case we can easily translate our three parameters into
physical ones and we have:
\begin{eqnarray}
c &=& 2 \tilde \alpha_0 A_0 \left[ 1+{\mu^2_0\over A_0}\right]^{3/2}\\
\tau &=& 2 A_0 \left(1+{\mu^2_0\over A_0}\right) \left[
\sqrt{1+{\mu^2_0\over A_0}}-1 \right]\\
\ln (\sigma) &=& 2 \left( 1+{\mu^2_0\over A_0}\right) 
\left[ {1\over\sqrt{1+{\mu^2_0\over A_0}}} -
{1\over\sqrt{1+{\mu^2_0+1\over A_0}}}\right] \nonumber\\
& & +\ln \left[4\left(A_0+\mu^2_0\right)\right]
+\ln\left[ {\sqrt{1+{1\over \mu^2_0+A_0}}-1\over 
\sqrt{1+{1\over \mu^2_0+A_0}}+1} \right]
\end{eqnarray}
where $\mu^2_0 = |m^2_{1,0}| M^2_{Pl}/V_0$ and $A_0 = 2\tilde m^2_0 
M^2_{Pl}/V_0$ are the values of the scalar and gaugino masses at
the Planck scale.
Then the allowed region for the $c, \sigma$ parameters shown in 
the first quadrant of Fig.~1 gives the bounds on 
physical parameters shown in Fig.~2.

\begin{figure}[t!] 
\centerline{
\epsfig{file=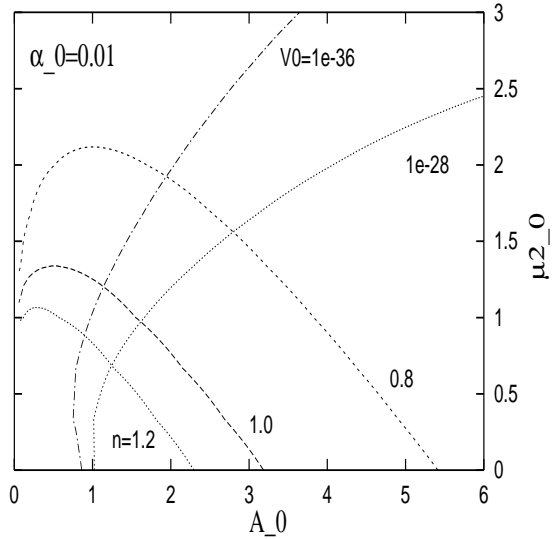,height=3in,width=3in}
}
\vspace{10pt}
\caption{ Lines of constant spectral index in the case of gauge 
dominated running of the inflaton mass in the plane 
$\mu^2_0 = |m^2_\phi(M_{Pl})| M^2_{Pl}/V_0$ vs 
$ A_0 = 2 \tilde m^2(M_{Pl}) M^2_{Pl}/V_0$ 
for a gauge coupling $\tilde \alpha_0 = N/(2\pi)
\,\alpha (M_{Pl}) = 0.01$. Also the lines of constant 
$V_0$ are displayed in units of $M^4_{Pl}$ for ${\cal N}_{COBE} = 45$.
This region corresponds to positive $\sigma$ and $c$ \cite{lcovi:cl98} .
}
\label{lcovi:2}
\end{figure}

Notice that in this particular case, an inflaton mass squared 
of order $V_0/M^2_{Pl}$ at the Planck scale is acceptable, and the 
running is efficiently flattening the potential, provided that the
gaugino mass is sufficiently large. Notice that, as shown in the graph
for a specific choice of the gauge coupling, generally gaugino
masses larger than the scalar one have to be assumed.

For consistency we have also to find the range of values of
$\lambda$ where this approximation is reliable: naturally the
limit $\lambda \rightarrow 0$ violates our assumption 
$\phi_{end} > \phi_c$, so that we have a lower bound on $\lambda$:
\begin{equation}
\lambda_0^2 \geq 4 V_0 \exp \left[ {2\over \tilde \alpha_0} 
\left( 1-{1\over \sqrt{1+{V_0+|m^2_{1,0}| \over A_0}} } \right) 
\right].
\label{boundlambda}
\end{equation}
As we can see this bound is very sensitive to the value of the gauge
coupling and also $V_0$; in general $\alpha_0$ has to be of the
order of $0.01$ or so to give a non negligible allowed region for the
initial masses and a non negligible allowed range for $\lambda$. 

\section{Dominant Yukawa coupling}

In this case the equations become similar to those for 
uncharged fields. We can therefore consider at the same time 
the model where $\phi_i$ are just two singlet fields substituting 
in the following $ N^2 \rightarrow 2$. This substitution
amounts to consider only one degree of freedom instead of the
$N^2-1$ of a field in the adjoint representation of $SU(N)$.
The solutions for the scalar masses are given by:
\begin{eqnarray}
m^2_S (t) &=& {N^2-1\over N^2+1} \left[ (m^2_{S,0} + m^2_{1,0} +
m^2_{2,0} + Y^2_0 ) {1\over 1-\tilde\lambda_0^2 t } \right. \nonumber  \\
& & \left. -Y_0^2 {1\over \sqrt{1-\tilde\lambda_0^2 t}} 
- m^2_{1,0} - m^2_{2,0} + {2\over N^2-1}
m^2_{S,0} \right] \\
m^2_i (t) &=& m^2_{i,0} + {1\over N^2-1} (m^2_S (t)-m^2_{S,0}),
\end{eqnarray}
where the subscript $0$ again indicates the initial values (defined 
at the Planck scale) and $\tilde \lambda^2_0 = {N^2+1 \over 2\pi^2} 
\lambda_0^2$.

We can see that in this case the initial conditions determine
whether one of the masses changes sign and an extremum in the 
potential is reached; generally the singlet mass appears to have
the stronger running since it interacts with more fields.
One possibility for inflation in this case, is to have negative 
initial masses and chose the inflaton to be the field whose
mass first become positive. 
In the case of universal initial masses such a field turns out 
to be the singlet; we have then that inflation happens in the
$S$ direction\footnote{Note that also the direction $S\neq 0,
\phi_i^a=0 $ is a flat direction of our potential.}
 and we can easily compute the parameters $c,\sigma,
\tau$ for vanishing $Y_0$:
\begin{eqnarray}
c &=& {4 (N^2-2)^2\over 3 N^4-1} \tilde\lambda_0\mu^2_0\\
\tau &=& {2 (N^2-2)\over 3 (N^2-1)} \mu^2_0\\
\ln (\sigma) &=& {1\over {2 (N^2-2)\over N^2+1} \mu^2_0-1}-
\ln \left[ 1 - {N^2+1\over 2 (N^2-2) \mu^2_0}\right],
\end{eqnarray}
where $\mu^2_0 = |m^2_S(M_{Pl})| M^2_{Pl}/V_0$.
In this case we have only two physical parameters to play with,
but an acceptable region exists as shown in Fig.~3: for initial
values of $\mu^2_0$ of order 1, a Yukawa coupling of order $0.05$
is needed to flatten the potential and $V_0$ is fixed by the
COBE normalization to be of order $10^{12-14} GeV$.

Notice that, in contrast with the previous case of dominant gauge
coupling, now $\mu^2_0$ plays both the role of the scalar
mass and of the gaugino mass and therefore a large initial
$\mu^2_0$ is needed in order for the running to be efficient.
As plotted in Fig. 3, $\mu^2_0$ has to be larger than $0.5$,
otherwise the $\eta$ parameter never becomes of order 1 and the
end of inflation has to be defined by the critical value 
$s_c$. In such a case the expression for $\sigma$ is much more
involved and we will not consider it.

\begin{figure}[t!] 
\centerline{
\epsfig{file=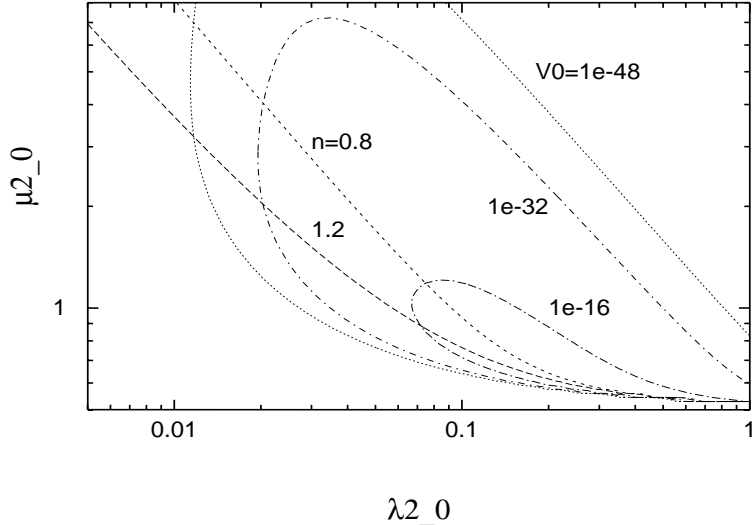,height=3in,width=4.5in}
}
\vspace{10pt}
\caption{ Lines of constant spectral index in the case of Yukawa 
dominated running of the inflaton mass in the plane 
$\mu^2_0 = |m^2_S(M_{Pl})| M^2_{Pl}/V_0$ vs 
$ \tilde \lambda^2_0$ for $N \gg 1$ and ${\cal N}_{COBE} = 45$. 
Also the lines of constant $V_0$ are displayed in units of $M^4_{Pl}$.
This region again corresponds to positive $\sigma$ and $c$.}
\label{lcovi:3}
\end{figure}

Inflation is also possible along the charged field direction,
but only for non universal masses and low values of $N$.
Taking as an example the case $N=2$, we have in terms of the
physical quantities, $\mu^2_0 = |m^2_{i,0}| M^2_{Pl}/V_0$
and $\xi^2_0 = m^2_{i,0}/m^2_{S,0}$:
\begin{eqnarray}
c &=& {(\xi^2_0-3)^2\over 5(\xi^2_0+2)} \tilde\lambda_0\mu^2_0\\
\tau &=& {\xi^2_0-3\over \xi^2_0+2} \mu^2_0\\
\ln (\sigma) &=& {1\over {\xi^2_0-3\over 5} \mu^2_0-1}-
\ln \left[ 1 - {5\over (\xi^2_0-3) \mu^2_0}\right].
\end{eqnarray}

In such case $\xi^2_0 > 3$, i.e. a singlet mass larger than the
charged fields mass, is needed for flattening the potential
(resembling the gaugino mass larger than scalar mass requirement
for the gauge dominated case).
The bounds on the physical parameter for this non universal case 
are given in Fig.~4.

\begin{figure}[t!] 
\centerline{
\epsfig{file=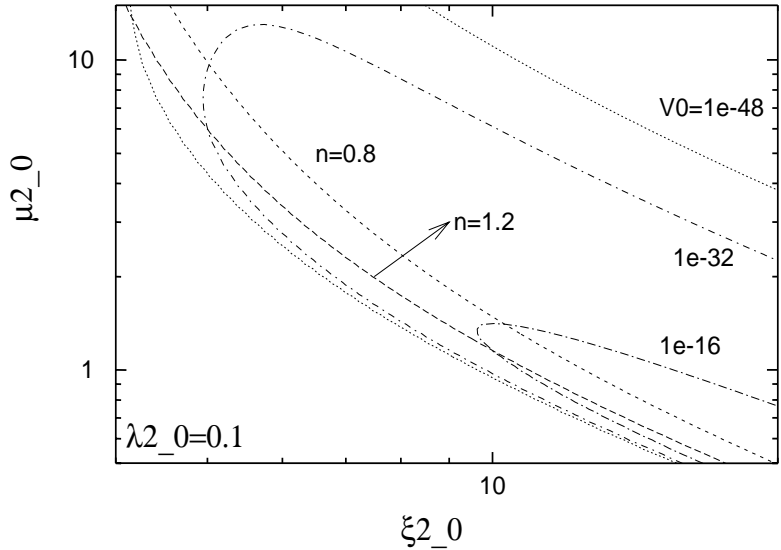,height=3in,width=4.5in}
}
\vspace{10pt}
\caption{ Lines of constant spectral index in the case of Yukawa 
dominated running of the inflaton mass for the non universal case.
It is shown the plane $\mu^2_0 = |m^2_{i,0}| M^2_{Pl}/V_0$ vs 
$ \xi^2_0 = m^2_{S,0}/m^2_{i,0}$ for $N =2$, $\tilde \lambda^2_0=0.1$
and ${\cal N}_{COBE} = 45$. Also the lines of constant 
$V_0$ are displayed in units of $M^4_{Pl}$.}
\label{lcovi:4}
\end{figure}

Another option is that of universal initial positive masses 
driven negative, or very small for what regards the inflaton, 
by the Yukawa coupling like in the case of the radiative EW 
breaking in the MSSM. In such a picture not only would the
quantum corrections be responsible for the flattening of the
potential, but also for the triggering of the hybrid--type 
end of inflation. Such a scenario would correspond to the
quadrants with negative $c$ in Fig.~1.

\section{Conclusions}

Quantum corrections can be strong enough to cancel the
supergravity contribution of order $V_0/M^2_{Pl}$ to the
inflaton mass and allow slow roll inflation to take place.
The requirement to have the spectral index in the experimental
range constraints tightly the parameter space of the specific 
models. Surprisingly anyway viable regions of the parameter
space exists for reasonable values of the couplings in the
different scenarios of gauge coupling or Yukawa coupling
dominance.

Since the running of the inflaton mass has to be substantial 
to give the cancelation, in this class of models the spectral 
index has a significant variation on cosmological scales, 
surely within the reach of the Planck satellite \cite{lcovi:planck}.
For example in the case of gauge coupling dominance the spectral
index changes by $0.1$ or so in the ten e-foldings corresponding 
to cosmological scales \cite{lcovi:clr98} and such a large 
variation could be observed or excluded even before the launch of
Planck, by the improvement of the data on the power spectrum of
density perturbations. The scale dependence of the spectral index
can be parameterized by eq. (\ref{lcovi:n-1}), assuming the linear 
approximation to be valid when cosmological scales left the 
horizon.

\section*{Acknowledgments}

I am very grateful and indebted to David H. Lyth and 
Leszek Roszkowski with whom this work has been done.
I would like to thank H. V. Klapdor-Kleingrothaus and 
the organizers of BEYOND 99 for the very interesting 
workshop and for financial support. 

This work was supported by PPARC grant GR/L40649.


\begin{thebibliography}{99}

\bibitem{lcovi:ly98} Lyth D H and Riotto A 1999 {\it
Phys. Rept.} {\bf 314} 1--146

\bibitem{lcovi:dine} Dine M, Fischler W and Srednicki M. 1981 
{\it Nucl. Phys.} B {\bf 189} 575 

\bibitem{lcovi:coughlan} Coughlan G D, Holman R, Ramond P and
Ross G G 1984 {\it Phys. Lett.} {\bf 140B} 44 

\bibitem{lcovi:cllsw} Copeland E J, Liddle A R, Lyth D H, 
Stewart E D and Wands D 1994 {\it Phys. Rev.} D {\bf 49} 6410 

\bibitem{lcovi:st97} Stewart E D 1997 {\it Phys. Lett.} {\bf 391B} 34

\bibitem{lcovi:st97bis}  Stewart E D 1997 {\it Phys. Rev.} D {\bf 56} 2019

\bibitem{lcovi:clr98} Covi L, Lyth D H and Roszkowski L 1999
{\it Phys. Rev.} D {\bf 60} 023509

\bibitem{lcovi:cl98} Covi L and Lyth D H 1999 {\it Phys. Rev.} D {\bf 59} 063515

\bibitem{lcovi:co98} Covi L 1999 {\it Phys. Rev.} D {\bf 60} 023513

\bibitem{ssb} D. Bailin and A. Love {\it Supersymmetric Gauge Field 
Theory and String Theory }, Istitute of Physics Publishing, Bristol 1994;

Nilles H P 1984 {\it Phys. Rep.} {\bf 110} 1--162

\bibitem{ma93} Martin S T and Vaughn M T 1994 {\it Phys. Rev.} D {\bf 50} 2282 

\bibitem{fd} Luty M A and Taylor W 1996 {\it Phys. Rev.} D {\bf 53} 3399 

\bibitem{lcovi:planck} home page at {\tt http://astro.estec.esa.nl/Planck}

\end{thebibliography}
\end{document}